%% file: GGAbsorption.tex
\documentclass[revtex4]{emulateapj}
\usepackage{graphicx}
\usepackage{amssymb}
\usepackage{amsmath}
\usepackage{ulem}
\usepackage{lipsum}
\usepackage[colorlinks=true,linkcolor=blue,anchorcolor=blue,citecolor=blue,urlcolor=blue]{hyperref}  

\usepackage{epstopdf}
\usepackage{layout}
\usepackage{subfigure}
\shorttitle{Gravitational Lensing to Avoid $\gamma - \gamma$ Absorption}
\shortauthors{Anna Barnacka, Markus B\"ottcher, \& Iurii Sushch}

\begin{document}

\title{How Gravitational Lensing Helps $\gamma$-ray Photons Avoid $\gamma - \gamma$ Absorption}

\author{Anna Barnacka\altaffilmark{1,2}, Markus B\"ottcher\altaffilmark{3,4} and Iurii Sushch \altaffilmark{3,5}}

\altaffiltext{1}{Harvard-Smithsonian Center for Astrohysics, Cambridge, MA 02138,USA}
\altaffiltext{2}{Astronomical Observatory, Jagiellonian University, Cracow, Poland}
\altaffiltext{3}{Centre for Space Research, North-West University, Potchefstroom, 2520, South Africa}
\altaffiltext{4}{Astrophysical Institute, Department of Physics and Astronomy, Ohio University, Athens, OH 45701, USA}
\altaffiltext{5}{Astronomical Observatory of Ivan Franko National University of L'viv, 79005, L'viv, Ukraine}

\email{Markus.Bottcher@nwu.ac.za, abarnacka@cfa.harvard.edu}

\begin{abstract}
We investigate potential $\gamma-\gamma$ absorption of $\gamma$-ray emission from blazars 
arising from inhomogeneities along the line of sight, beyond the diffuse Extragalactic Background 
Light (EBL). As plausible  sources of excess $\gamma-\gamma$ opacity, we consider (1) foreground 
galaxies, including cases in which this configuration leads to strong gravitational lensing, (2) 
individual stars within these foreground galaxies, and (3) individual stars within our own galaxy, 
which may act as lenses for microlensing events. We found that intervening galaxies close to the 
line-of-sight are unlikely to lead to significant excess $\gamma-\gamma$ absorption. This opens up 
the prospect of detecting lensed gamma-ray blazars at energies above 10~GeV with their gamma-ray 
spectra effectively only affected by the EBL. The most luminous stars located either in intervening 
galaxy or in our galaxy provides an environment in which these gamma-rays could, in principle, be 
significantly absorbed. However, despite a large microlensing probability due to stars located in 
intervening galaxies, $\gamma$-rays avoid absorption by being deflected by the gravitational 
potentials of such intervening stars to projected distances (``impact parameters'') where the 
resulting $\gamma-\gamma$ opacities are negligible. Thus, neither of the intervening excess photon
fields considered here, provide a substantial source of excess $\gamma-\gamma$ opacity beyond
the EBL, even in the case of very close alignments between the background blazar and a foreground
star or galaxy. 
\end{abstract}

\keywords{galaxies: active - galaxies: jets - gamma-rays: gravitational lensing - strong, micro}

\section{ Introduction}
\input{introduction}

\section{\label{absorption}Gamma-ray absorption}
\label{sec:absorption}
\input{absorption}

\section{\label{stronglenses}Intervening Galaxies}
\label{sec:stronglenses}
\input{stronglenses}

\section{\label{microlensesCosmological}Microlensing Stars within the Intervening Galaxy}

\input{microlensesCosmological}

\label{sec:microlensesCosmological}

\section{\label{microlenses}Microlensing Stars within the Milky Way}
\label{sec:microlenses}
\input{microlenses}

\section{ Summary}
\label{sec:summary}
\input{summary}

\section*{Acknowledgments}
The work of A.B. is supported by the Department of Energy Office of Science, NASA \& the
Smithsonian Astrophysical Observatory.
The work of M.B. is supported through the South African Research Chair Initiative (SARChI)
by the National Research Foundation and the Department of Science and Technology of South
Africa, under SARChI Chair grant No. 64789.

\bibliography{GGAbsorption}
\end{document}

%% file: introduction.tex
The extragalactic $\gamma$-ray sky is dominated by blazars, 
which are a class of radio-load active galactic nuclei (AGNs) 
with relativistic jets viewed at small angles with respect to the jet axis. 
The radiation of blazars is dominated by non-thermal emissions from the jets. 
The spectral energy distribution of blazars is characterized by two broad components.
The low-energy (radio through UV or X-rays) component is produced by the synchrotron 
radiation of relativistic electrons. The origin of the high-energy (X-rays through
$\gamma$-rays) component is still under debate, and both leptonic and hadronic scenarios
are viable. In leptonic models, the X-ray through $\gamma$-ray emission is the result 
of inverse-Compton radiation with seed photons originating from within the jet (i.e.,
the synchrotron radiation), or external to the jet, such as from the broad-line region 
or a dusty torus. In hadronic emission models, the $\gamma$-ray emission results from
proton synchrotron radiation and photo-pion induced cascade processes \citep[for a 
comprehensive discussion of leptonic and hadronic emission models, see, e.g.,][]{Boettcher13}. 

The blazar class is divided into two sub-classes based on the presence or absence of 
optical and UV emission lines, which are likely correlated with the strengh of external 
photon fields in the blazar environment. Objects exhibiting prominent emission lines
have historically been classified as flat spectrum radio quasars (FSRQs) and the presence 
of a broad line region implies a significant external radiation field. In the second 
sub-class historically classified as BL Lac objects, only weak (Equivalent width $\le
5$~\AA) or no emission lines are typically detected, which provides no evidence for
the presence of a substantial broad-line region and, hence, for a strong external
radiation field. 

$\gamma$-rays with energies above a few tens of GeV, produced in relativistic jets, 
may be substantially affected by interactions with various photon fields, which cause 
$\gamma-\gamma$ absorption and electron-positron pair production. The interaction of 
$\gamma$-ray emission with the thermal radiation  from the dusty torus, the broad line 
region or the accretion disc may produce an imprint in the spectra of blazars \citep{2003APh....18..377D}.
When the emission region in which $\gamma$-rays are produced (often called the ``blazar zone'') 
is located within the broad line region, the $\gamma$-rays have to pass through this
intense radiation field dominated by Ly$\alpha$ line emission. As a result, $\gamma-\gamma$ 
absorption may produce a break in the observed spectra of FSRQs \citep{2010ApJ...717L.118P}.
The $\gamma -\gamma$ absorption effect is also used to probe central region of AGNs 
\citep{2011ApJ...728..134R,2010ApJ...717..468R} and to constrain the location of the 
$\gamma$-ray emitting region \citep{2013arXiv1307.1779B}. 

$\gamma$-ray observations of blazars at energies above 100~GeV are precluded for sources 
located at large redshift ($z \gtrsim 1$) because of $\gamma-\gamma$ absorption by the Extragalactic 
Background Light (EBL). In addition to the EBL, overdensities of target photons for $\gamma-\gamma$
absorption may arise if a galaxy is located close to the line-of-sight between the blazar and 
the observer. When the center of the intervening galaxy is located at a projected distance of a few 
kpc from the line-of-sight, the emission may be split into several paths and can be magnified by
strong gravitational lensing.

At high-energy $\gamma$-rays (HE:~100~MeV$<$E$<$100~GeV), two strongly gravitationally lensed 
blazars have been observed thus far \citep{2011A&A...528L...3B,2014ApJ...782L..14C}. As has been 
recently proposed by \citet{2014arXiv1403.5316B}, gravitationally lensed blazars offers a way 
to investigate the structure of the jet at high energies and, thus, to locate the site of the
$\gamma$-ray production within the jet. $\gamma$-ray emission from the jets cannot be spatially 
resolved due to limited angular resolution of current detectors, thus the observations of strongly 
gravitationally lensed blazars are very valuable in order to investigate the origin of $\gamma$-ray 
emission of blazars. However, $\gamma$-rays produced within relativistic jets of gravitationally lensed 
blazars have to pass through, or at least in close proximity to, the intervening galaxy on their way 
to the observer. One might therefore plausibly expect that the infrared -- ultraviolet radiation fields 
of galaxies (or other intervening matter) acting as lenses may lead to excess $\gamma-\gamma$ absorption
of the blazar $\gamma$-ray emission. 

In this paper, we investigate whether observations of gravitationally lensed blazars at energies above 
10~GeV are not precluded by $\gamma-\gamma$ absorpion. To this aim, in section~\ref{sec:absorption}, we
provide a general introduction to the $\gamma-\gamma$ absorption process and a simple estimate of the
$\gamma-\gamma$ opacity produced by near-line-of-sight light sources in a point-source approximation.
In section~\ref{sec:stronglenses}, we consider the probability of having an intervening galaxy sufficiently
close to the line-of-sight to a background blazar to cause an observable gravitational-lensing effect, and 
we calculate the $\gamma-\gamma$ opacity due to such a lensing galaxy under realistic assumptions concerning
the luminosity, spectrum, and radial brightness profile of the galaxy. In section \ref{sec:microlensesCosmological} 
we investigate microlensing and $\gamma$-ray absorption effects by stars within the intervening galaxy. 
In section \ref{sec:microlenses} we consider the chance of observing microlensing effects due to stars
within our own Galaxy in the light curves of blazar observed by {\it Fermi}/LAT. We summarize in 
section~\ref{sec:summary}.

%% file: absorption.tex
Gamma-ray photons emitted from sources at cosmological distances may be subject to 
$\gamma-\gamma$ absorption as they travel through various photon fields. 
The universe is transparent for gamma-ray photons with energy below $\sim$10~GeV.
Above these energies, the gamma-ray horizon is limited due to absorption by interactions 
with infrared -- ultraviolet radiation fields, composed of integrated light emitted by 
stars and infrared emissions reprocessed by dust, accumulated throughout the history
of the Universe. These low-energy photon fields are known as the Extragalactic Background 
Light (EBL), and are the subject of extensive studies 
\citep{2006ApJ...648..774S,2006Natur.440.1018A,2007ApJ...666..663B,2008A&A...487..837F,2010ApJ...712..238F,2010ApJ...723.1082A,2012Sci...338.1190A}.

The fundamental process responsible for $\gamma-\gamma$ absorption is the interaction 
with low-energy photons via electron-positron pair-production \citep{1967PhRv..155.1404G}:
\begin{equation}
\gamma_{HE} + \gamma_{LE} \rightarrow e^+ + e^- \,.
\label{reaction}
\end{equation}

The threshold condition for the pair production is:
\begin{equation}
\epsilon_{HE} \epsilon_{LE} \, (1-\cos\theta) > 2
\label{threshold}
\end{equation}
where $\epsilon = h \nu / (m_e c^2)$ denotes the normalized photon energy, and
$\theta$ is the interaction angle between the gamma-ray and the low-energy
photon. The $\gamma-\gamma$ absorption cross section has a distinct maximum at
$\sim$~twice the threshold energy. Therefore, TeV photons will primarily be
absorbed by infrared radiation, with wavelength $\lambda_{LE}$ estimated by 

\begin{equation}
\lambda_{LE} = 2.4 \, E_{TeV} \, \mu{\rm m}
\label{lambdaLE}
\end{equation}

The optical depth for photon-photon absorption, $\tau_{\gamma \gamma}$, is
given by \citep{1967PhRv..155.1404G}:

\begin{equation}
\tau_{\gamma \gamma} (\epsilon_{HE}) = \int \mbox{d}l \int \mbox{d}\Omega \, (1-\mu) \int 
\mbox{d} \epsilon \, n(\epsilon, \Omega; l) \, \sigma_{\gamma\gamma} \, (\epsilon_{HE}, \epsilon, \mu)
\label{taugg}
\end{equation}
where d$l$ is differential path traveled by the $\gamma$-ray photon, $d\Omega = d\phi \, d\mu$,
$\mu = \cos\theta$, $n(\epsilon, \Omega; l)$ is the low energy photon number density, and the
$\gamma-\gamma$ absorption cross section is given by \citep{JR76}

%\begin{widetext}
%\begin{equation}
$$\sigma_{\gamma\gamma} (\epsilon_1, \epsilon_2, \mu) = {3 \over 16} \, \sigma_T \, (1 - \beta_{\rm cm}^2)
$$
\begin{equation}
\cdot \left( [3 - \beta_{\rm cm}^4] \, \ln\left[ {1 + \beta_{\rm cm} \over 1 - \beta_{\rm cm}} \right]
- 2 \beta_{\rm cm} \, [2 - \beta_{\rm cm}^2] \right)
\label{sigmagg}
\end{equation}
%\end{widetext}
where $\beta_{\rm cm} = \sqrt{1 - 2 / (\epsilon_1 \, \epsilon_2 \, [1 - \mu])}$ is the normalized
velocity of the newly created electron and positron in the center-of-momentum frame of the 
$\gamma-\gamma$ absorption interaction.

In addition to diffuse EBL, gamma-rays may suffer from substantial absorption when an intense 
source of light is close to the line-of-sight between the source and the observer. Such an intensive 
source of light may be provided by a foreground galaxy or stars within it, or a star within our own
galaxy. For a point source with a narrow (e.g., thermal) photon spectrum peaking at characteristic
photon energy $\epsilon_s$, the differential photon density may be approximated by

\begin{equation}
n(\epsilon_s, \Omega; l) = {L \over 4 \pi \, x^2 \, c\, \epsilon_s^2 \, m_e c^2 } \, 
\delta(\Omega - \Omega_s)
\label{npointsource}
\end{equation}
where $L$ is the total luminosity of the source, $x$ is the distance between the source and any given 
point along the gamma-ray path, and $\Omega_s$ describes the solid angle in the direction 
towards the source from that point. For a simple estimate, we use a $\delta$ function approximation 
to the $\gamma-\gamma$ absorption cross section, $\sigma_{\gamma\gamma} (\epsilon_1 \, \epsilon_2) 
\approx (\sigma_T/3) \, \epsilon_1 \, \delta(\epsilon_1 - 2/\epsilon_2)$. If the impact parameter of
the $\gamma$-ray path (i.e., the distance of closest approach to the source) is $b$, and we define
$l = 0$ as the point of closest approach, then $x = \sqrt{b^2 + l^2}$ and $\mu = l/x$. With these
simplification, the integration in Equ. (\ref{taugg}) can be evaluated analytically to yield an 
estimate for the $\gamma-\gamma$ optical depth at a characteristic $\gamma$-ray energy of $E_{\gamma}
= 520 \, E_{eV}^{-1}$~GeV (where $E_{eV}$ is the target photon energy in units of eV):

\begin{equation}
\tau_{\gamma \gamma} (E_{\gamma}) \approx {\sigma_T \, L \over 6 \, c \, m_e c^2 \, \epsilon_s \, b}
= 3 \times 10^{-9} \left( {L \over L_{\odot}} \right) \, E_{eV}^{-1} \, b_{\rm pc}^{-1}
\label{tau_point}
\end{equation}
where $b_{\rm pc}$ is $b$ in units of pc. Conversely, we can use Equ. (\ref{tau_point}) to define a
``$\gamma-\gamma$ absorption sphere'' of radius $r_{\rm abs}$ within which the line of sight would 
need to pass a source for $\gamma$-rays to experience substantial ($\tau_{\gamma\gamma} > 1$) 
$\gamma-\gamma$ absorption:

\begin{equation}
r_{\rm abs} \sim 10^9 \, \left( {L \over L_{\odot}} \right) \, E_{eV}^{-1} \;
{\rm cm} \sim 73  \, \left( {L \over L_{\ast}} \right) \, E_{eV}^{-1} \; {\rm pc}
\label{rabs}
\end{equation}
where the latter estimate assumes a characteristic luminosity of of a galaxy, $L_{\ast} = 2.4 \times 10^{10}
\, L_{\odot}$. Equ. (\ref{rabs}) suggests that only the most massive stars are capable of causing
significant $\gamma-\gamma$ absorption individually, which will be confirmed with more detailed
calculations in Section \ref{sec:microlensesCosmological}. For entire galaxies, typically the
$\gamma-\gamma$ absorption sphere, as evaluated by Equ. (\ref{rabs}) is smaller than the 
galaxy itself, which means that the line of sight would need to pass through the galaxy (in
which case, of course, our approximation of the galaxy as a point source becomes invalid). 
The case of $\gamma-\gamma$ absorption by intervening galaxies will be considered with more
detailed calculations in Section \ref{sec:stronglenses}.

%% file: stronglenses.tex
Let us now consider the effect of an intervening galaxy close to the line-of-sight between an observer 
and a blazar. When the projected distance between the lens and the source, in the lens plane, is of the 
order of a few kpc or less, strong lensing phenomena are expected. A critical parameter determining lensing
effects is the Einstein radius, defined as: 

\begin{eqnarray}
r_E &=& \theta_E\times D_{OL}
=\sqrt{\frac{4GM}{c^2} D } \nonumber  \\
&\approx &
5\,\mbox{kpc} \, \left(
\frac{D}{1\,\mbox{Gpc}} \right)^{\frac{1}{2}} \left( \frac{M}{10^{11}
  \,\mbox{M}_{\odot}} \right)^{\frac{1}{2}}
\,,
\label{eq:thetaE}
\end{eqnarray} 

where $D=D_{OL} D_{LS}/ D_{OS}$, $D_{OL}$, $D_{LS}$, and $D_{OS}$ are the angular diameter distances from the observer to the lens, 
from the lens to the source and from the observer to the source, respectively 
\citep{1996astro.ph..6001N,1992grle.book.....S,2013arXiv1307.4050B}, and we have scaled the
expression to the typical mass of a lensing galaxy, $\sim 10^{11}\,\mbox{M}_{\odot}$. 

The chance of lensing by an object like a galaxy or a star within the galaxy can be expressed in 
terms of the lensing optical depth, $\tau_L$, which is a measure of the probability that at any 
instance in time a lens is within an angle $n \times \theta_E$ of a source:

\begin{equation}
\tau_L (D_{OS}) = \frac{\pi n^2}{d\Omega} \int \mbox{d} V_L \int \mbox{d} M \, \rho_L (M, D_{OL}) \theta^2_E (M,D) \,,
\end{equation}

where $n$ denotes the number of Einstein radii in which we are looking for a potential lens,
$\mbox{d}V_L = \mbox{d}\Omega D_{OL}^2 \mbox{d}D_{OL}$ is a differential volume element on a shell 
with radius $D_{OL}$ covering a solid angle $d\Omega$, and $\rho_L (M, D_{OL})$ is the number density 
of potential lenses \citep{1992grle.book.....S}. 

\begin{figure}[ht]
\includegraphics[width=6.2cm,angle=-90]{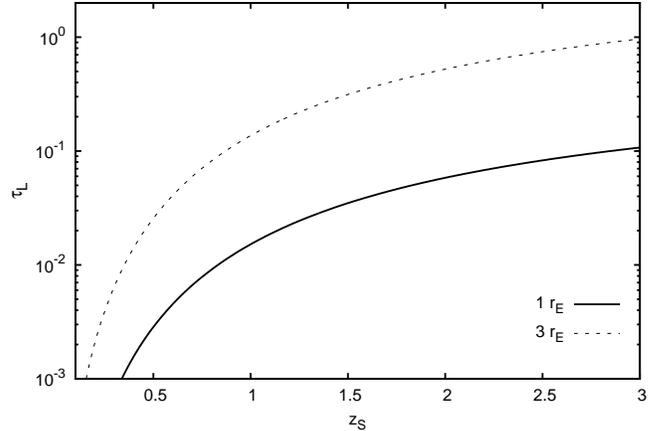}%
\caption{\label{fig:Tau} Total lensing optical depth, $\tau_L$, 
as a function of the redshift of the source, $z_S$. The curves show 
the optical depth accounting for the lenses within $1 \, r_E$ (solid) 
and within 3~$r_E$ (dashed).}
\end{figure}

Figure~\ref{fig:Tau} shows the total optical depth as a function of the redshift of the source.  
The calculations are based on a homogenous Friedmann-Lema{\^i}tre-Robertson-Walker cosmology,
with Hubble constant $H_0$=67.3~km/s/Mpc, mean mass density $\Omega_M=0.315$ and the normalized 
cosmological constant $\Omega_\Lambda=0.686$ \citep{2013arXiv1303.5076P}. When the lensing optical 
depth is small, it can be interpreted as the probability of a background galaxy being located 
within $n$ Einstein radii of the center of mass of a foreground galaxy.

\citet{2013arXiv1307.4050B} has estimated a number of expected strongly, gravitationally-lensed 
systems among 370 FSRQs listed in the 2nd Fermi catalogue \citep{2012ApJS..199...31N}.
In the sample of these 370~FSRQs, the expected number of sources with at least one intervening 
galaxy within one Einstein radius was estimated to be $\sim10$.

Therefore, on average, 3~\% of FSRQs detected by the {\it Fermi}/LAT have a galaxy located within a 
projected distance smaller than one Einstein radius ($\sim$5~kpc). Extrapolating this result to the distance 
of $3 \times r_E$, one can expect that on average $\sim 30$~\% of gamma-ray blazars will have an 
intervening galaxy within a distance smaller than 15~kpc, i.e. within $3 \times r_E$. For any given 
blazar, the number of foreground galaxies, within a certain distance from the line-of-sight, 
depends on the redshift of the source (see Figure~\ref{fig:Tau}). 

The projected distance, or impact parameter, $b$, in the foreground galaxy plane, is defined as 
the distance between the center of mass of a foreground galaxy and the line-of-sight between 
the emitting region of the source and the observer. When the system satisfies the condition that 
$b$ is larger than $r_E$, one is in the weak lensing regime \citep{2005astro.ph..9252S,2001PhR...340..291B}.
In this regime, the morphology of the image is slightly deformed and may be displaced, but in general 
there is only one image of the source, lensing magnification is negligible, and therefore the light 
curve of a source remains unaffected.

On the contrary, when $b$ is smaller than $r_E$, the light is split into several paths, and the
images may be significantly magnified. The position and magnification of images change with $b$.
When mass distribution of a lens is well represented by a singular isothermal sphere (SIS), 
there are two images. A third image has zero flux, and can therefore be ignored.

The positions of the images are at $r_{\pm} = b \pm r_E$ \citep{1996astro.ph..6001N}.
When $b$ approaches $r_E$, the $r_+$ image appears beyond $r_E$; at the same time, $r_-$ moves 
toward the center of the lens. The light of the image at $r_-$ will take a path closer to the 
center of the galaxy, where one could suspect that $\gamma-\gamma$ absorption in the radiation 
field of the lensing galaxy may be non-negligible. 

The magnification of lensed images is given as $A_{\pm} = r_\pm / b$. When image, $r_-$, appears
closer to the center of the lens, its magnification decreases, so that emission from these images 
becomes negligible in the limit of very small distances from the center. On the contrary, the second 
image is further deflected from the center of the lens, and is strongly magnified. Therefore, lensed
images with significant magnification will pass the galaxy at large distances where $\gamma-\gamma$
absorption might be negligible. 

In order to estimate whether $\gamma-\gamma$ absorption by a lensing galaxy might be important, the
corresponding opacity $\tau_{\gamma\gamma}$ needs to be evaluated. As pointed out in Section 
\ref{sec:absorption}, the point source approximation used to derive the estimate in Equ. (\ref{tau_point})
is not valid for impact parameters of the order of (or smaller than) the effective radius of the galaxy. 
We therefore, evaluate the integral in Equ. (\ref{taugg}) numerically, properly accounting for the 
angular dependence of the extended radiation field of the galaxy $n(\epsilon, \Omega; l)$. For this
purpose, we approximate the galaxy as a flat disk with a De Vaucouleurs surface brightness profile:

\begin{equation}
F(r) = F_0 \, e^{-a \, \left( \left[ {r \over r_e} \right]^{1/4} - 1 \right)}
\label{Devaucouleurs}
\end{equation}
where $a = 3.33$, $r_e$ is the effective radius of the galactic bulge, and $F_0$ can be related to
the total luminosity $L$ of the galaxy through $F_0 \approx 2.14 \times 10^{-3} \, L / r_e^2$. The
spectrum of the galaxy is approximated by a blackbody radiation field with temperature $T = \Theta
\, m_e c^2 / k$, so that the spectral disk flux as is represented as 

\begin{equation}
F_{\epsilon} (r) = K \, e^{-a \, \left( \left[ {r \over r_e} \right]^{1/4} - 1 \right)} \; 
{\epsilon^3 \over e^{\epsilon/\Theta} - 1} 
\label{diskspectrum}
\end{equation}
with $K = \pi^4 \, F_0 / (15 \, \Theta^4)$. 

\begin{figure}[ht]
\includegraphics[width=10cm]{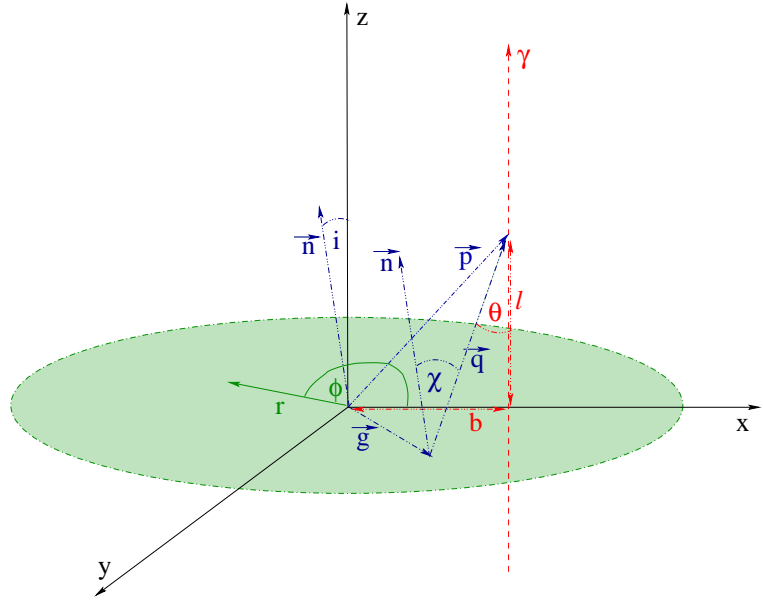}
\caption{\label{galaxy_geometry} Geometry used for the calculation of the $\gamma-\gamma$ opacity due
to the radiation field of a lensing galaxy.}
\end{figure}

Fig. \ref{galaxy_geometry} illustrates the geometry adopted for the calculation of $\tau_{\gamma\gamma}$:
We chose the direction of propagation of the $\gamma$-ray as the $z$ axis, and the $x$ axis is defined
by the radius vector from the center of the galaxy to the impact point of the $\gamma$-ray trajectory
on the galactic disk, at distance $b$ (the impact parameter). The normal vector of the galactic disk
is inclined with respect to the $z$ axis by the inclination angle $i$. The integration over the galactic
disk is carried out in polar co-ordinates with the angle $\phi$ measured from the $x$ axis. The photon
propagation length $l = 0$ is defined as the photon's impact point on the disk. 
$\overrightarrow g$ denotes the radial vector from the center of the galaxy to any given point $(r, \phi)$
on the disk, $\overrightarrow p$ is the vector from the center to the current location of the
$\gamma$-ray photon, and $\overrightarrow q = \overrightarrow p - \overrightarrow g$ is the vector
connecting any point on the disk to the location of the $\gamma$-ray photon, whose length can be
calculated as

\begin{equation}
q = \sqrt{ r^2 + l^2 + b^2 - 2 \, r \, b \, \cos\phi - 2 \, r \, l \, \sin\phi \, \sin i}
\label{q_galaxy}
\end{equation}

The $\gamma\gamma$ interaction angle cosine is then given by $\mu = \cos\theta = q_z / q = (l - r \, \sin\phi \, 
\sin i)/q$. Finally, we note that the solid angle element $d\Omega$ can be re-written by considering the projected 
disk surface element: $q^2 \, d\Omega = \vert \cos\chi \vert \, r \, dr \, d\phi$, where $\chi$ is the angle between 
$\overrightarrow q$ and the disk normal, $\overrightarrow n = (0, \, \sin i, \, \cos i)$:

\begin{equation}
\cos\chi = {2 \, r \, \sin\phi \, \cos i \, \sin i - l \, \cos i \over q}
\label{coschi}
\end{equation}
which allows us to replace the $d\Omega$ integration by an integration over the disk surface, $r \, dr \, d\phi$. 
Thus, the $\gamma\gamma$ opacity is calculated as

%\begin{widetext}
%\begin{equation}
$$\tau_{\gamma\gamma} (\epsilon_{\rm HE}) = {K \over 4 \, \pi \, m_e \, c^3} \, \int\limits_{-\infty}^{\infty} d l
\int\limits_r^{\infty} r \, dr \, e^{- a \left( \left[ {r \over r_e} \right]^{1/4} - 1 \right)}
$$
\begin{equation}
\cdot \int\limits_0^{2\pi}
d\phi \, (1 - \mu) \, {\vert \cos\chi \vert \over q^2} \, \int\limits_{2 \over \epsilon_{\rm HE} \, (1 - \mu)}^{\infty} 
d\epsilon \, {\epsilon^2 \over e^{\epsilon/\Theta} - 1} \, \sigma_{\gamma\gamma} (\epsilon_{\rm HE}, \epsilon, \mu). 
\label{taugg_galaxy}
\end{equation}
%\end{widetext}

\begin{figure}[ht]
\vskip 1cm
\begin{center}
\includegraphics[width=6.2cm,angle=-90]{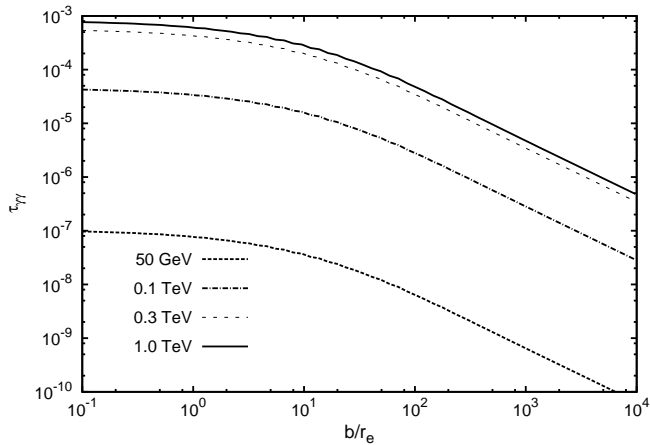}
\end{center}
\caption{\label{taugg_galaxy_re07} $\gamma -\gamma$ opacity as a function of impact parameter $b$ of the
(assumed undeflected) $\gamma$-ray path from the center of the galaxy, for various $\gamma$-ray energies.
The galaxy is assumed to be a Milky-Way like galaxy with an effective radius of $r_e = 0.7$~kpc, intercepted 
at an inclination angle of $i = 30^o$, a temperate $T=6000$~K and a luminosity $L=L_*$. }
\end{figure}

Figure \ref{taugg_galaxy_re07} shows the resulting $\gamma-\gamma$ opacity for gamma-ray photons
passing through a Milky-Way like galaxy ($L = L_{\ast}$, $r_e = 0.7$~kpc), as a function of the impact
parameter $b$, for various gamma-ray energies. The results are shown for an inclination angle of $i =
30^o$, but we find that they are only very weakly dependent on $i$. The figure illustrates that for such
a case, $\gamma-\gamma$ absorption within the collective radiation field of an individual, intervening 
galaxy is negligible, irrespective of the gamma-ray's impact parameter. We note that for $b \gg r_e$, 
where the galaxy may reasonably be approximated by a point source, our results are in excellent agreement 
with the analytical estimate of Equ. (\ref{tau_point}). 

Obviously, the $\gamma-\gamma$ opacity scales linearly with the galaxy's luminosity, so given the same
radial profile (with $r_e = 0.7$~kpc), a luminosity of $L \gtrsim 100 \, L_{\ast}$ would be required
to cause significant $\gamma -\gamma$ absorption. 

\begin{figure}[ht]
\vskip 1cm
\begin{center}
\includegraphics[width=6.2cm,angle=-90]{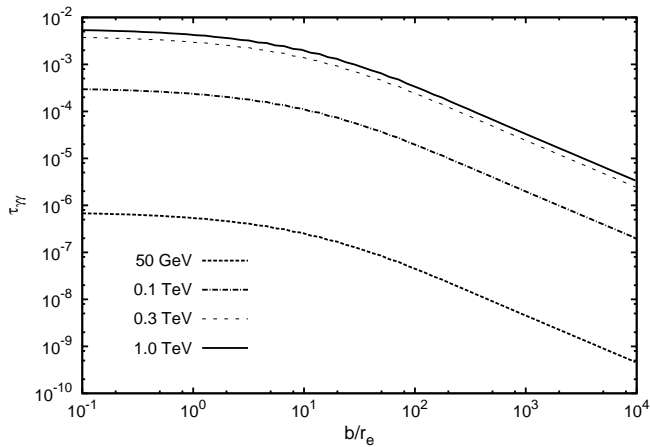}
\end{center}
\caption{\label{taugg_galaxy_re01} Same as Figure \ref{taugg_galaxy_re07}, but for an effective radius 
of $r_e = 0.1$~kpc. }
\end{figure}

Figure~\ref{taugg_galaxy_re01} shows the $\gamma-\gamma$ opacity due to an $L_{\ast}$ galaxy, assuming an
effective radius of $r = 0.1$~kpc. As expected, the maximum opacity (for small impact parameters) increases 
with increasing compactness of the galaxy. However, even for this case, an extreme luminosity of $L \gtrsim
10 L_{\ast}$ (combined with very small size) would be required to lead to substantial $\gamma-\gamma$ absorption.
However, galaxies of such luminosities are typically giant ellipticals or large spirals with substantially 
larger effective radii than 1~kpc. We may therefore conclude that, in any realistic lensing situation, 
$\gamma-\gamma$ absorption due to the collective radiation field of the galaxy is expected to be negligible.

%% file: microlensesCosmological.tex
When gamma-rays emitted by a source at a cosmological distance (e.g., a blazar)
crosses the plane of an intervening galaxy, they pass a region with an over density 
of stars and therefore have a non-negligible probability of passing near the 
$\gamma-\gamma$ absorption sphere of a star, as estimated by Equ. (\ref{rabs}). This suggests
that they may suffer non-negligible $\gamma-\gamma$ absorption. 

For a more detailed evaluation of the $\gamma-\gamma$ opacity as a function of
$\gamma$-ray photon energy $\epsilon_{HE}$ and impact parameter $b$, we have
evaluated $\tau_{\gamma\gamma} (\epsilon_{HE})$ numerically. This is done by
numerically carrying out the integrations in Equ. (\ref{taugg}) under a point
source approximation, as discussed in Section \ref{sec:absorption}, representing 
the stellar spectrum as a blackbody with characteristic temperature and luminosity
determined by the spectral type of the star, and using the full $\gamma-\gamma$
absorption cross section (Equ. \ref{sigmagg}). 

\begin{figure}[ht]
\includegraphics[width=6.2cm,angle=-90]{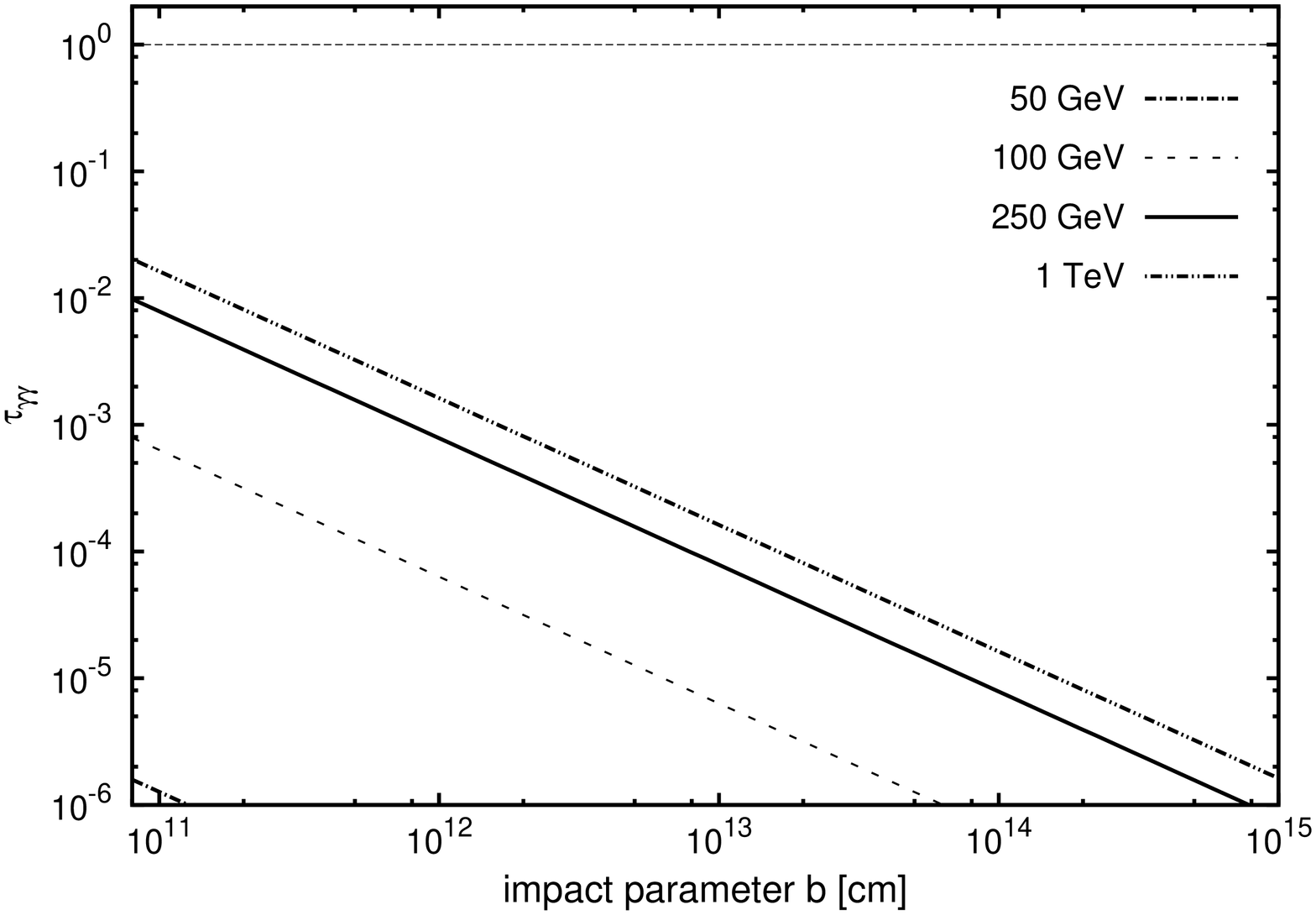}
\hskip 0.2cm
\includegraphics[width=6.2cm,angle=-90]{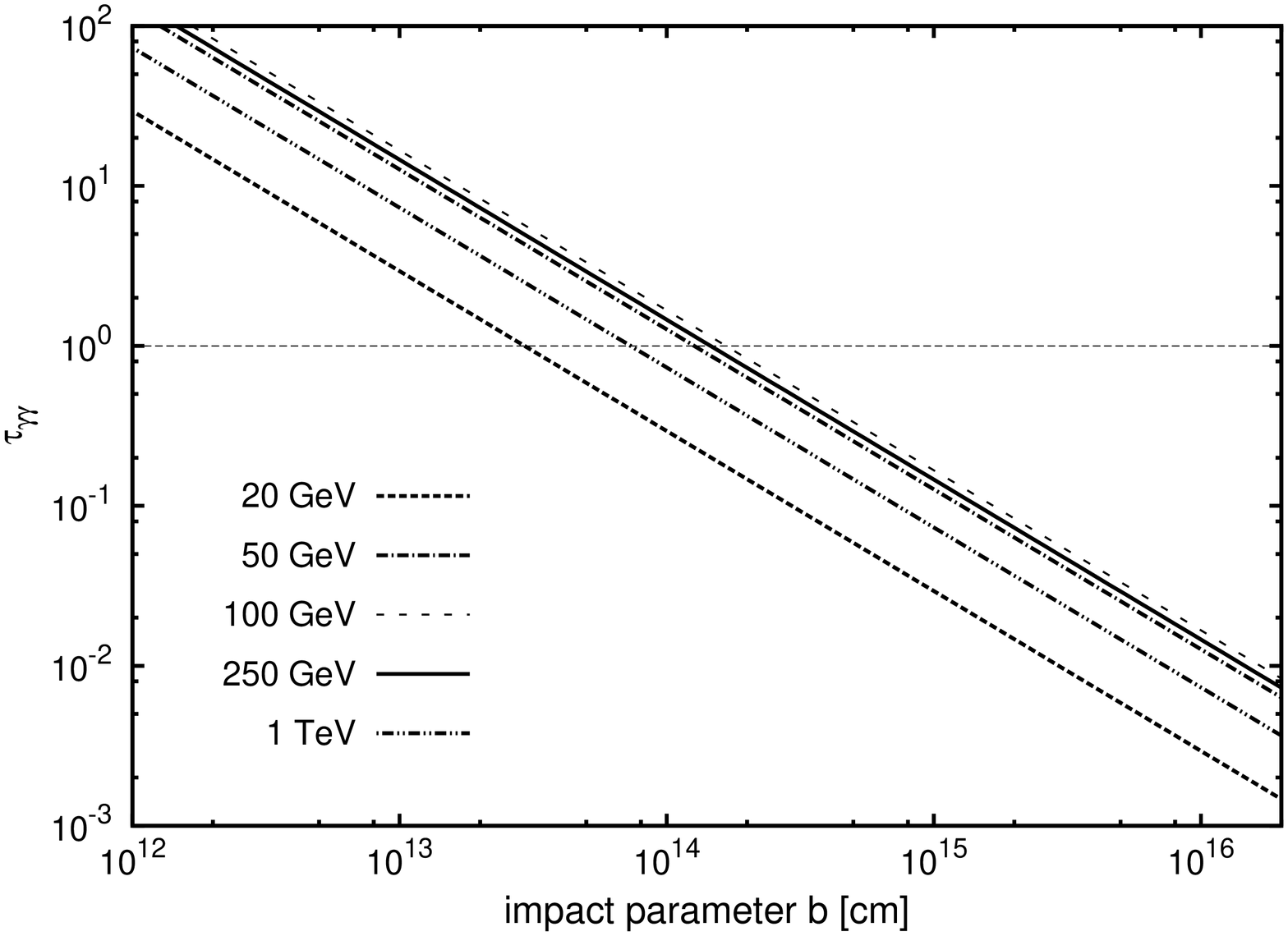}
\caption{\label{fig:taugg} The optical depth for $\gamma-\gamma$ absorption, $\tau{\gamma\gamma}
(E_{\gamma})$ as a function of the impact parameter, measured from the centers of star. Stellar 
luminosities, blackbody temperatures, and radii are chosen corresponding to a G2-type main-sequence 
star (left) and an O5-type main-sequence star (right). All curves start at the stellar radius and 
are labeled by the respective $\gamma$-ray energy, $E_{\gamma}$. }  
\end{figure}

Figure~\ref{fig:taugg} shows the optical depth for $\gamma-\gamma$ absorption as a 
function of the impact parameter, i.e., distance of closest approach to the center 
of the star, for various $\gamma$-ray photon energies, for a sun-like
(G2V) and a very massive (O5V) star. In both figures, the curves begin at the radius
of the star. The figure illustrates that for a sun-like star, no significant $\gamma-\gamma$
absorption is expected for any line of sight that does not pass through the star. We
find that the same conclusion holds for all stars with spectral type F0 or later
(i.e., less massive and cooler). For A-type stars, $\gamma-\gamma$ absorption (though 
still with maximum $\tau_{\gamma\gamma} < 1$) can occur if the line of sight passes 
within a few stellar radii from the surface of the star. Significant $\gamma-\gamma$ 
absorption (with $\tau_{\gamma\gamma}^{\rm max} > 1$) can occur for more massive (O and 
B) stars, within a few tens to hundreds of stellar radii. The right panel of 
Fig. \ref{fig:taugg} illustrates that significant $\gamma-\gamma$ absorption 
by an O-type star can occur for impact parameters up to $\sim 10^{15}$~cm from 
the center of the star. We note that the results of our numerical calculations 
are in excellent agreement with the back-of-the-envelope estimate provided in 
Equ. (\ref{tau_point}).

The calculations presented above assume straight photon paths, unaffected by gravitational lensing. 
The deflection of the light-rays by the stars in the foreground lens galaxy, so-called microlensing 
effect, has been widely elaborated \citep{1986ApJ...301..503P,1986A&A...166...36K,1989AJ.....98.1989I,2007ASPC..371...43K}.
Given the small deflection angles expected from microlensing, our calculations of $\tau_{\gamma\gamma}$
are still expected to be accurate, as long as a proper value for the impact parameter $b$ is used that
takes into account the microlensing effect. It is known that the probability of microlensing of 
gravitationally-lensed quasars by the stars in the foreground lens galaxy is 1 \citep{2006astro.ph..4278W}, 
which means that $\gamma$-rays are likely to pass through the Einstein radius of many stars before 
escaping from the galaxy. 

The Einstein radius of a star located at cosmological distances is of the order of $r_E \sim 
5 \times 10^{16} \, (M/M_{\odot})^{1/2}$~cm (see Equ. \ref{eq:thetaE}). The radii of main-sequence 
stars range from $\sim 10^{10}$ -- $10^{12}$~cm, which is much smaller than $r_E$. Therefore, the 
mass distribution of the lens is well approximated by a point-mass. 

\begin{figure}[h!]
\includegraphics[width=6.2cm,angle=-90]{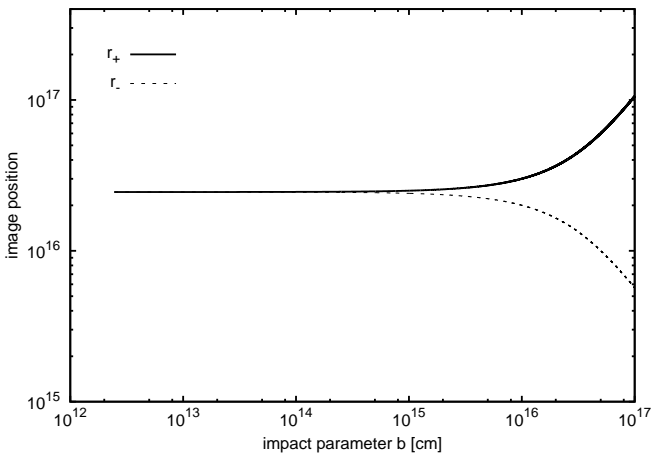}%
\caption{\label{fig:ratioES} The distance of the lensed images, $r_+$ and $r_-$, from the center of the 
lense as a function of the projected distance $r_S$ between the source and the lens. The source is at 
redshift $z_S = 0.6$, and lens at redshift $z_L = 0.4$. The lens is assumed to have a mass of $M_{\rm lens}
= 1 \, M_{\odot}$.}  
\end{figure} 

\begin{figure}[h!]
\includegraphics[width=6.2cm,angle=-90]{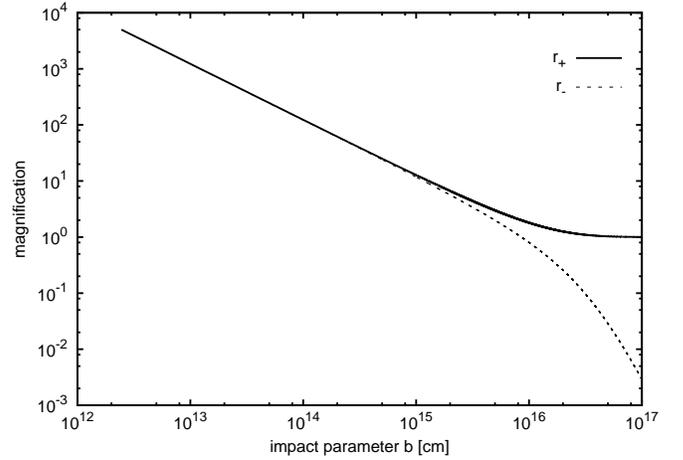}
\caption{\label{fig:MicroMagnification} Magnification ofthe  lensed images as 
a function of the projected distance between the source and the lens, $r_S$,
in the lens plane, for the same parameters as used in Figure \ref{fig:ratioES}. }  
\end{figure}

Figure \ref{fig:ratioES} shows the positions $r_{\pm}$ of the lensed images of a background
source, as a function of the distance between a source and a lens, $b$, in the lens plane. Both
the source and the lens have been assumed to be point-like, resulting in two source images at
positions $r_{\pm}$. The lens has a mass of $M_{\rm lens} = 1 \, M_{\odot}$. The figure 
illustrates that for very small projected distances $b$ both images appear at $r_{\pm} \sim 
2.5 \times 10^{16}$~cm, which implies that both lines of sight pass the star far outside the 
$\gamma$-ray absorption sphere. Only for projected distances near the Einstein radius does 
one of the images (at $r_-$) appear at smaller distances from the star. 
Figure~\ref{fig:MicroMagnification} shows the magnification of the lensed images for this
case. This figure shows that, when the $r_-$ image appears at small separations from the
lens, it will be strongly de-magnified and, thus, any modulation of this image by $\gamma-\gamma$
absorption will remain undetectable, while the outer, essentially un-magnified image will remain
unaffected by $\gamma-\gamma$ absorption due to its large impact parameter from the star. 

An additional factor playing into the consideration of potential lensing and
$\gamma\gamma$ absorption effects, is the apparent size of the $\gamma$-ray source on
the plane of the sky. The sizes of $\gamma$-ray sources can generally be constrained based
on causality arguments, from the minimum variability time scale, which typically yields
sizes of the order of $10^{16}$~cm \citep{2011AdSpR..48..998S} at the distance of the source. 
The size of such a source projected onto the lens planes is $D_{OS}/D_{OL}$ times larger.
Thus, the projected size of the $\gamma$-ray emitting region is comparable to the size of 
the Einstein radius of a solar-mass sized lens. Therefore, the source will appear extended
in the plane of the lens, and will probe a multitude of sight-lines around the lens
but no matter which sight line any individual $\gamma$-ray photon will 
follow, gravitational lensing will always cause the photon to avoid the $\gamma$-ray 
absorption sphere of the lens. 

As a result, gamma-rays which travel cosmological distances and pass through galaxies with 
over-densities of stars, will not suffer substantial $\gamma-\gamma$ absorption because 
$\gamma$-rays will be deflected to the distances far beyond the $\gamma$-ray absorption
radius around any individual star in the intervening galaxy. 

Light traces matter, with the relation between sources of light and of space-time
curvature being determined by the mass-to-light ratio $M/L$ of any given source. Thus, our
considerations suggest that for any sources of light with similar (or larger) $M/L$ as the
sources considered here (i.e., stars or galaxies), gravitational lensing will always aid in 
avoiding $\gamma-\gamma$ absorption by intervening light sources at cosmological distances.

%% file: microlenses.tex
We finally consider the microlensing effects in the light curves of blazars produced 
by stars in our own Galaxy. In this case the Einstein radius of solar mass stars is 
of the order of a few $\times 10^{13} \, (M / M_{\odot})^{1/2}$~cm, i.e., smaller by 
a factor of $\sim 1000$ compared to cosmological lenses, due to their smaller distance. 
This has an impact on the lensing optical depth, $\tau$, which for  Galactic microlensing 
is of the order of $10^{-6}$. 
The duration of a microlensing event is given by the time required for the lens to move
by $2r_E$ relative to the line-of-sight to the background source. With
typical Galactic speeds of $v \sim 200$km~s$^{-1}$, the time scale of typical Galactic 
microlensing events  is $t_0 \sim 130$~days~$\times (M/M_\odot)^{1/2}$. 

For the purpose of a rough estimate, let us assume that all lensing objects have the same 
mass and the same velocity. Then the number of microlensing events, $N$, that may be 
expected if $n_S$ sources are monitored over a time interval $\Delta t$, can then be estimated
as

\begin{equation}
N=\frac{2}{\pi} \, n_S \, \tau \frac{\Delta t}{t_0} \,.
\end{equation}

The {\it Fermi} satellite has monitored the entire sky since August 2008, corresponding to
$\Delta t \sim2000$~days. The observations performed over the first five years of its operation 
resulted in the detection of $\sim 1000$ objects. Using these estimates, the number of expected 
microlening events in the light curves of objects monitored by {\it Fermi}/LAT is of the order 
of $10^{-3}$. Therefore, microlensing effects by stars located in our Galaxy are extremely rare.
The $\gamma\gamma$ absorption radius of a star in our Galaxy is still substantially smaller than 
$r_e$ for low-mass stars, and comparable to $r_e$ for the most massive stars. The same conclusion
therefore holds for potential $\gamma\gamma$ absorption effects from intervening stars in our
Galaxy, so that $\gamma$-rays can travel freely through our Galaxy without lensing deflection and/or 
$\gamma\gamma$ absorption by stars.

%% file: summary.tex
Blazars are the most luminous (non-transient) sources detected up to large cosmological distances.
The lensing probability for these luminous and distant sources is thus significant:
of the order of a few percents for sources observed in the energy range from 100~MeV to 300~GeV.
The gravitational lensing by intervening galaxies and individual stars
within these lensing galaxies may lead to repeating $\gamma$-ray
light curve patterns due to the time delay between the lensed images of the blazar.
The measurement of these time delays and magnification ratios between the flaring episodes of the given blazars
can be exploited to ascertain the location of the $\gamma$-ray emitting regions 
within the blazars \citep{2014arXiv1403.5316B}. 

In this paper, we have investigated whether the light emitted by
the foreground lenses may affect the lensing signatures due to $\gamma-\gamma$ absorption.
We found that the collective photon fields from lensing galaxies are not expected to
produce any measurable excess $\gamma-\gamma$ opacity beyond that of the EBL, 
and that microlensing and $\gamma-\gamma$ absorption by stars within our own galaxy 
is extremely unlikely to affect any {\it Fermi}/LAT detected $\gamma$-ray blazars. 
Our most intriguing result is that microlensing stars within intervening galaxies
are not expected to lead to significant excess $\gamma-\gamma$ absorption either,
as the gravitational lensing effect will always cause the light paths reaching
the Earth to be deflected around the source of excess light, keeping them at distances
from the star, at which $\gamma-\gamma$ absorption remains negligible. 

Consequently, we have demonstrated that light curve studies of gravitationally-lensed
blazars are a promising avenue for revealing the structure of the $\gamma$-ray emitting 
region in blazars, and that excess $\gamma-\gamma$ absorption by the radiation fields 
of gravitational lenses will not interfere with $\gamma$-ray lensing studies. 
As such, the magnification ratio between echo flares in the light curve
is not affected by the $\gamma-\gamma$ absorption and future observations 
of blazars at energies above 10~GeV with experiments like VERITAS \citep{2002APh....17..221W}
or CTA \citep{2011ExA....32..193A} are not precluded by $\gamma$-ray absorption.